\documentclass[aps,pre,
 onecolumn,
12pt,
 notitlepage,
 showpacs,floatfix,nofootinbib,
 double-space
 superscriptaddress
 ]{revtex4-1}
\usepackage[english]{babel}
\usepackage{subcaption}
\usepackage{bbold}
\usepackage{amssymb}
\usepackage{amsfonts}
\usepackage{amsmath}
\usepackage{amsthm}
\usepackage{epsfig}
\usepackage{float}
\usepackage{multirow}
\usepackage[usenames,dvipsnames]{color}
\usepackage[latin1]{inputenc}
\usepackage{xr}
\usepackage{enumitem}
\usepackage{microtype}
\usepackage{graphics,graphicx,xcolor}
\usepackage{comment}
\usepackage{cancel}
\usepackage{lipsum}

\newcommand{\av}[1]{\left \langle #1 \right \rangle}
\newcommand{\sig}[2]{\vec{\sigma}_{#1}^{#2}}
\newcommand{\kk}[1]{\mathbf{K}_{#1}}

\newcommand{\thx}[0]{\theta^{*}}

\graphicspath{{newfigs/}}

\begin{document}

\title{Multistability and Juggling States in Frustrated Kuramoto Oscillators on Ring Networks}

\author{Guilherme S. Costa$^{1}$}
\author{Ricardo Fariello$^2$}
\author{Marcus A.~M.~de Aguiar$^{1}$}

\affiliation{$^1$Instituto de F\'isica Gleb Wataghin, Universidade Estadual de Campinas, 13083-970, Campinas, SP, Brazil}
\affiliation{$^2$ Departamento de Ci\^encias da Computa\c{c}\~ao, Universidade Estadual de Montes Claros, 39401-089, Montes Claros, MG, Brazil}

\begin{abstract}

We study the properties of matrix coupled oscillators in ring networks and show that the system supports several families of synchronized clustered solutions. One of these families is characterized by consecutive oscillators belonging to distinct clusters, called alternate ordering, with two classes of solutions. The first consists of one or two static clusters, that lock on phases that depend on the coupling parameters, despite the non-zero natural oscillation frequency. The second one is comprised of clusters that rotate rigidly. The number of clusters depends on the neighbor connectivity. We also investigated sequential ordered clusters and showed that the constraints imposed by the interface oscillators imposes additional restrictions to equilibrium configurations. Although some of these solutions are stable, their basins of attraction depend on model parameters: when the system is initialized with random phases, the oscillators tend to organize themselves into dynamical clusters whose interfaces move along the ring. The oscillators within each cluster remain close to one of the stable phase configurations, while particles are exchanged at the interfaces. When interfaces moving in opposite directions collide, the corresponding clusters annihilate, until a state with interfaces moving in a common direction is reached.

\end{abstract}

\maketitle

\section{INTRODUCTION}

The synchronization of coupled oscillators plays a fundamental role in many physical, chemical, and biological systems \cite{pikovsky2001synchronization, boccaletti2018}. Examples include fireflies \cite{buck1976synchronous}, cardiac pacemaker cells \cite{Michaels704}, the collective firing of neurons \cite{gray1994}, chaotic systems \cite{boccaletti2002synchronization}, and metronomes on a platform \cite{Pantaleone2002}. The wide variety of systems exhibiting similar transitions to synchrony led Winfree to propose a general formulation to describe weak interactions between nonlinear oscillators of arbitrary nature \cite{Winfree1967,Strogatz2000}. 

Winfree reasoned that oscillatory systems should have a stable limit cycle responsible for its periodic pattern. Moreover, the phase of system along the cycle should be the key variable describing its dynamics. Cycle stability would also ensure that changes in the amplitude of oscillation could be discarded even when small perturbations were introduced to the system. In Winfree's model, a set of $N$ oscillators are coupled according to  
\begin{equation}
	\dot{\theta_i} = \omega_i + Q(\theta_i) \sum_{j=1}^{N}P(\theta_j)
\end{equation}  
where $\omega_i$'s are natural frequencies, accounting for the fact that oscillators are not necessarily identical, $P(\theta_j)$ is the {\it pulse function} produced by oscillator $j$ and $Q(\theta_i)$ is the response of oscillator $i$ to the pulse field generated by the other oscillators. Different functional forms of $P$ and $Q$ were recently studied in the context of specific problems \cite{Pazo2014,Okeeffe2016,Chandra2017,Gallego2017}. 

Kuramoto showed later that Winfree's equations could also be written as  \cite{kuramoto1975} 
\begin{equation}
	\label{eq::kuramotoOriginal}
	\dot{\theta_i} = \omega_i + \dfrac{K}{N} \sum_{j=1}^{N} \operatorname{sin}(\theta_j - \theta_i),
\end{equation}
where the sinusoidal coupling is the simplest case of a more general class of coupling functions involving phase differences \cite{Strogatz2000}. 

Winfree and Kuramoto models can be unified and generalized if the oscillators are described by unit vectors $\vec{\sigma}_i=(\cos\theta_i,\sin\theta_i)$ \cite{chandra2019} and the coupling constant is promoted to a coupling matrix acting on these vectors \cite{Buzanello2022}. Coupling the oscillators by a matrix can also be interpreted as a form of generalized frustration \cite{Aguiar2023}. Different coupling matrices lead to different types of frustration, as we review in the next section. The Kuramoto equations, in particular, are obtained if the coupling matrix is proportional to the identity. One important property of the model is the possibility of breaking the rotation symmetry, a feature of the Winfree model that is lost in the Kuramoto equations  \cite{Chandrasekar2020,mihara2025rotational}. Several aspects of the Kuramoto model with generalized frustration were recently explored, including its application to modular networks \cite{COSTA2026110410}, forced oscillators and Arnold tongues \cite{costa2026arnold} and higher dimensional oscillators \cite{Aguiar2023,Fariello2024}. However, in the majority of these previous studies, the oscillators were assumed to be connected all-to-all, which facilitates theoretical approaches. 

In this paper we study the properties of matrix coupled oscillators in ring networks. The ring topology is particularly interesting because it can represent a set of spatially separated units where interaction is possible only between sufficiently close units. Ring networks of Kuramoto oscillators have been studied in different contexts in the past few years, including time-delayed systems \cite{ameli2025synchronization}, phase locking and multistability on cell complexes \cite{bavcic2026phase}, unidirectional coupling \cite{ha2012basin}, and synchronization in  2-simplicial ring networks \cite{wang2026synchronization}. Rings of identical Kuramoto oscillators with nearest-neighbor coupling were also studied in detail in Ref. \cite{denes2019pattern}. 

We show that rings of matrix coupled Kuramoto oscillators exhibit several equilibrium solutions in which the oscillators group themselves into synchronized clusters. However, starting from random phases, the typical asymptotic solution involves the formation of dynamical clusters: although the phase within each cluster remains constant, the particles belonging to them change over time. The clusters exchange boundary particles similar to a juggling performance. The number of possible dynamical clusters and the probability of each specific solution, which is a measure of its basin of attraction, depends on the coupling parameters and the connectivity of the ring. The set of oscillators belonging to the clusters moves along the ring with constant speed, similar to solitons \cite{mclaughlin1978perturbation}. Collision of two such sets leads to their annihilation and the asymptotic state corresponds to clusters moving in the same direction with the same speed, avoiding further collisions.

The rest of this paper is divided as follow: In Section \ref{sec:model}, we describe the frustrated, or matrix coupled, Kuramoto model, discussing previous results and the novelties related to this model. In Section \ref{sec:analytical}, we look for equilibrium solutions consisting of static or rigid rotated clusters, tailored from specific initial conditions while in Section \ref{sec:general} we investigate the general dynamics of the model. Finally, in Section \ref{sec:discussion}, we draw our final remarks.

\section{MODEL}
\label{sec:model}

Kuramoto oscillators can be described by their phases $\theta_i$ or, alternatively, by unit vectors $\vec{\sigma_i} = (\cos \theta_i,\sin \theta_i)$. In terms of vectors, Eq. (\ref{eq::kuramotoOriginal}) becomes
\begin{equation}
	\label{eq::vecform}
	\dfrac{d \sig{i}{}}{dt} = \textbf{W}_i \sig{i}{} + \dfrac{K}{N} \sum_{j=1}^{N} [ \sig{j}{} - (\sig{i}{}\cdot \sig{j}{}) \sig{i}{}],
\end{equation}
in which 
\begin{equation}
	\textbf{W}_i = 
	\begin{pmatrix} 
		0 & \omega_i \\
		-\omega_i &0 
	\end{pmatrix}
\end{equation}
is the frequency matrix. Although the vector equations are completely equivalent to the phases equations, the former can be easily generalized to higher dimensions by simply extending $\vec\sigma_i$ to a $D-$dimensional unit vector, described by $D-1$ spherical angles and moving on the surface of a unit $D-$sphere. The only requirement is that frequency matrix $\textbf{W}_i$ must be anti-symmetric \cite{chandra2019}.

In this formulation, the order parameter can be calculated as the average of the vectors $\sig{i}{}$
\begin{equation}
	\vec{r} = \dfrac{1}{N}\sum_{i=1}^{N} \vec{\sigma}_i  = (r \sin{\psi}, r\cos\psi).
	\label{eq::orderP}
\end{equation}
Another important feature of the vector formulation is that the coupling constant $K$ can be naturally extended to a $D \times D$ real matrix $\kk{}$ \cite{Buzanello2022}, that acts on the vectors and modifies the dynamics according to
\begin{equation}
	\dfrac{d \sig{i}{}}{dt} = \textbf{W}_i \sig{i}{} + \dfrac{1}{N} \sum_{j=1}^{N} [ \mathbf{K}\sig{j}{} - (\sig{i}{}\cdot \mathbf{K} \sig{j}{}) \sig{i}{}].
	\label{eq::kuraFrust}
\end{equation}

The coupling matrix plays the role of generalized frustration, in which the vector $\sig{j}{}$ is rotated and scaled by $\kk{}$ before interacting with $\sig{i}{}$. Using the definition in Eq.(\ref{eq::orderP}), the dynamical equation can also be written as
\begin{equation}
	\dfrac{d \sig{i}{}}{dt} = \textbf{W}_i \sig{i}{} + [ \mathbf{K}\vec{r} - (\sig{i}{}\cdot \mathbf{K} \vec{r}) \sig{i}{}].
	\label{eq::kuraFrustMF}
\end{equation}

Several versions of the Kuramoto model can be obtained as particular cases of Eq.\eqref{eq::kuraFrust} by fixing $D=2$ and choosing the matrix $\kk{}$ accordingly \cite{COSTA2026110410}. For example, if $\kk{}$ is proportional to a rotation matrix
\begin{equation}
	\textbf{K} =  K 
	\begin{pmatrix}
		\cos \alpha       & \sin \alpha  \\
		-\sin \alpha & \cos \alpha 
	\end{pmatrix},
\end{equation}
the equation for the phases corresponds to the Kuramoto-Sakaguchi model \cite{Sakaguchi1986}
\begin{equation}
	\dot{\theta}_i = \omega_i + \frac{1}{N} \sum_{j=1}^{N} K \sin (\theta_j - \theta_i - \alpha)
\end{equation}
which further reduces to the original Kuramoto model when $\alpha=0$. For
\begin{equation}
	\textbf{K}  =  
	\begin{pmatrix}
		0 & 0  \\
		0 & \lambda
	\end{pmatrix}
\end{equation}
the equation for the phases is of the type proposed by Winfree \cite{Winfree1967}, 
\begin{equation}
	\dot{\theta_i} = \omega_i + \lambda \cos\theta_i \sum_{j=1}^{N} \sin{\theta_j}
\end{equation}  
where $\cos\theta$ is the response function and $\sin\theta$ is the pulse function. Notice that, in this case, rotational symmetry is broken, as $\kk{}$ has real eigenvalues and a single real eigenvector in the y-direction, that provides a special direction in the plane. 

For $D=2$, which is the case of interest in this work, the coupling matrix $\kk{}$ can be conveniently parameterized as \cite{Buzanello2022}
\begin{equation}
	\textbf{K} \equiv \kk{R} + \kk{S} =  K 
	\begin{pmatrix}
		\cos \alpha       & \sin \alpha  \\
		-\sin \alpha & \cos \alpha 
	\end{pmatrix}
	+ J
	\begin{pmatrix}
		-\cos \beta       & \sin \beta  \\
		\sin \beta & \cos \beta 
	\end{pmatrix},
	\label{matrizK}
\end{equation}
reducing the equations for the phases $\theta_i$ to the more familiar form:
\begin{equation}
	\dot{\theta}_i = \omega_i + \sum_{j=1}^{N} \left[ K \sin (\theta_j - \theta_i - \alpha) + J \sin(\theta_j + \theta_i + \beta)\right].
	\label{eq::thetai}
\end{equation}

This equation also arises in the context of Stuart-Landau limit-cycle oscillators with conjugate feedback \cite{Chandrasekar2020}. For $J$ and $\alpha$ equals to zero, $\kk{}$ is proportional to the identity, leading to the usual Kuramoto model. For $K=0$ and $J>0$, on the other hand, identical oscillators with $\omega_i=0$ exhibit full synchronization with zero angular velocity at $\theta_i=-\beta/2$. These are termed {\it phase tuned} states, as the phase of the cluster can be controlled by $\beta$, which points in the direction of the dominant eigenvector of $\kk{}$. In the general case where $K$ and $J$ are both non-zero, the system exhibits as active states, where the phase and module of $\vec{r}$ oscillate in time. As rotational symmetry is generally broken, the value of $\omega_0$, the center of distribution of natural frequencies $g(\omega)$, plays a direct role in the dynamics \cite{Buzanello2022}.

\section{ANALYTICAL SOLUTIONS FOR RING NETWORKS}
\label{sec:analytical}

The model introduced in the previous section assumed all-to-all interactions. 
If, instead, the oscillators are connected by a network with adjacency matrix $A$, the frustrated model becomes
\begin{equation}
	\dot{\theta}_i = \omega_i + \dfrac{1}{k_i}\sum_{j=1}^{N} A_{ij} \left[ K \sin (\theta_j - \theta_i - \alpha) + J \sin(\theta_j + \theta_i + \beta)\right]
	\label{eq::thetainets}
\end{equation}
where $A_{ij}=1$ if oscillators $i$ and $j$ interact and zero otherwise, and $k_i=\sum_j A_{ij}$ is the degree of $i$. For the case of a ring network with $2Q-$neighbors the dynamical equations for identical oscillators become
\begin{align}
\dot{\theta}_i
=\omega+\frac{K}{2Q}\sum_{n=1}^{Q}
\left[\sin\left(\theta_{i+n}-\theta_i + \alpha \right)
+\sin\left(\theta_{i-n}-\theta_i+ \alpha \right) \nonumber \right]\\
+\frac{J}{2Q}\sum_{n=1}^{Q} 
\left[\sin\left(\theta_{i+n}+\theta_i + \beta \right)
+\sin\left(\theta_{i-n}+\theta_i + \beta \right)
\right].
\end{align}

\subsection{Fully synchronized solutions}

Let us first investigate the fully synchronized solution $\theta_i = \thx, \forall i$. The equations reduce to 
\begin{align}
\dot{\theta}
=\omega+K\sin \alpha+ J\sin(2\thx+\beta),
\label{eq::homo}
\end{align}
and have equilibrium solutions if $|J| \geq |\omega + K\sin{\alpha}|$. These solutions can be separated in two families:

\begin{align}
\thx_1 &= \frac{1}{2} \left[ \arcsin\left(-\frac{\omega+K\sin\alpha}{J}\right)
-\beta\right]+n\pi \qquad \text{and} \nonumber \\
\thx_2 &= \frac{1}{2} \left[\pi -\arcsin\left(-\frac{\omega+K\sin\alpha}{J}\right)
-\beta\right]+n\pi,
\label{eq::solutions}
\end{align}
in which $n\in\mathbb{Z}$. Restricting the phases in the interval $[0,2\pi]$, each of the families generate two solutions, for a total of $4$ equilibrium phases. When $|J| = |\omega + K\sin{\alpha}|$ the two families coalesce leading to only $2$ distinct stationary solutions. To study the stability of these solutions we perform a small perturbation in the dynamical equations and obtain
\begin{align}
\delta\dot{\theta}_i=&\frac{K\cos\alpha+J\cos(2\theta^*+\beta)}{2Q}
\sum_{n=1}^{Q}\left[\delta\theta_{i+n}+\delta\theta_{i-n}
\right]\nonumber\\
&+\left[-K\cos\alpha+J\cos(2\theta^*+\beta)\right]\delta\theta_i.
\end{align}
The Jacobian matrix, therefore, is circulant and the eigenvalues are given by
\begin{equation}
	\lambda_m = \sum_{k=1}^N a_k e^{2\pi i m (k-1)/N}
    \label{eq::eigen}
\end{equation}
where $a_k$ are the entries in the first line of the matrix. Defining
\begin{align}
R_p^{(Q)}
=
\frac{1}{Q}
\sum_{k=1}^{Q}
\cos\left(\frac{2\pi pk}{N}\right),
\end{align}
we obtain
\begin{align}
\lambda_p
=
J\cos(2\theta^*+\beta)
\left(1+R_p^{(Q)}\right)
-
K\cos\alpha
\left(1-R_p^{(Q)}\right).
\end{align}

For $p=0$, we have $R_0^{(Q)}=1$, and therefore
\begin{align}
\lambda_0 = 2J\cos(2\theta^*+\beta).
\end{align}
This mode corresponds to a homogeneous perturbation recovering the stability condition obtained from Equation \eqref{eq::homo}. In this case the family $\thx_1$ is unstable and $\thx_2$ is stable for $J > 0$. If $J < 0$, the families switch their stabilities.

Using the equilibrium condition,
\begin{align}
\sin(2\theta^*+\beta)
=
-\frac{\omega+K\sin\alpha}{J},
\end{align}
the equilibrium branch satisfies
\begin{align}
J\cos(2\theta^*+\beta)
=
\pm\sqrt{J^2-(\omega+K\sin\alpha)^2}.
\end{align}
Consequently, the eigenvalues associated with the stable synchronized
branch are
\begin{align}
\lambda_p
=
-\sqrt{J^2-(\omega+K\sin\alpha)^2}
\left(1+R_p^{(Q)}\right)
-
K\cos\alpha
\left(1-R_p^{(Q)}\right).
\label{eq:lam}
\end{align}

The fully synchronized state is therefore linearly stable if
\begin{align}
\max_{p\neq0}\lambda_p < 0.
\end{align}

For $p\neq0$, the function $R_p^{(Q)}$ can also be written as
\begin{align}
R_p^{(Q)}
=
\frac{
\sin\left(\frac{\pi pQ}{N}\right)
\cos\left(\frac{\pi p(Q+1)}{N}\right)
}{
Q\sin\left(\frac{\pi p}{N}\right)
}.
\end{align}

In particular, when $K\cos\alpha>0$, both terms in Eq.(\ref{eq:lam}) are non-positive, 
since $-1\leq R_p^{(Q)}\leq1$. Thus, in this case,
the stability of the homogeneous mode $p=0$ implies stability against
spatial perturbations.

For $K\cos\alpha<0$, the two terms compete, and the stability condition
for each non-homogeneous mode can be written as
\begin{align}
\sqrt{J^2-(\omega+K\sin\alpha)^2}
>
|K\cos\alpha|
\frac{1-R_p^{(Q)}}{1+R_p^{(Q)}}.
\end{align}
The most restrictive condition is obtained from the mode that maximizes $\lambda_p$.

In addition to the fully synchronized solution, we investigated two additional cases comprising of equally distributed $m$-cluster solutions, namely alternate ordering and sequential orderingsynchronization. In the former, we look for solutions in which consecutive oscillators belong to distinct clusters. In the latter, all oscillators belonging to a cluster have consecutive labels. For example, for $10$ oscillators divided in $2$ clusters ($C_1$ and $C_2$), in the alternate scheme the distribution is $\{C_1,C_2,C_1,C_2,C_1,C_2,C_1,C_2,C_1,C_2\}$ while the sequential distribution is
$\{C_1,C_1,C_1,C_1,C_1,C_2,C_2,C_2,C_2,C_2\}$.

\subsection{Alternate ordering - $2Q$ neighbors}

Let us assume that the oscillators are grouped into $m$ synchronized clusters and that the phase of all oscillators in cluster $l$ is $2\pi l/m$. Consider the equations for the oscillators of a generic cluster $l$:
\begin{align}
\dot{\theta}_l
=
\omega
+
\frac{K}{2Q}
\sum_{n=1}^{Q}
\left[
\sin\left(
\theta_{l+n}-\theta_l+\alpha
\right)
+
\sin\left(
\theta_{l-n}-\theta_l+\alpha
\right)
\right]
\nonumber\\
+
\frac{J}{2Q}
\sum_{n=1}^{Q}
\left[
\sin\left(
\theta_{l+n}+\theta_l+\beta
\right)
+
\sin\left(
\theta_{l-n}+\theta_l+\beta
\right)
\right].
\end{align}
Since
$\theta_{l\pm n}-\theta_l=\pm 2\pi n/m$, the $K$ terms can be
combined using the identity
$\sin(a+b)+\sin(a-b)=2\sin(a)\cos(b)$.
In addition, considering that
\begin{align}
\theta_{l\pm n}+\theta_l
=
2\gamma+
\frac{4\pi(l-1)}{m}
\pm
\frac{2\pi n}{m},
\end{align}
the equation can be written as
\begin{align}
\dot{\theta}_l
=
\omega
+
\frac{K\sin\alpha}{Q}
\sum_{n=1}^{Q}
\cos\left(\frac{2\pi n}{m}\right)
+
\frac{J}{Q}
\sin\left(
2\gamma+
\frac{4\pi(l-1)}{m}
+\beta
\right)
\sum_{n=1}^{Q}
\cos\left(\frac{2\pi n}{m}\right).
\end{align}

By defining
\begin{align}
F(Q,m)
=
\sum_{n=1}^{Q}
\cos\left(\frac{2\pi n}{m}\right),
\end{align}
the conditions for the existence of $m$ cluster solutions are that either
$F=0$ or the sine term is independent of $l$. Let us investigate
each of them separately. Using the Dirichlet kernel
\begin{align}
1+2\sum_{n=1}^{Q}\cos(nx)
=
\frac{\sin((Q+1/2)x)}{\sin(x/2)},
\end{align}
with $x=2\pi/m$, the condition $F=0$ simplifies to
\begin{align}
\sin\left(
\left(Q+\frac{1}{2}\right)
\frac{2\pi}{m}
\right)
=
\sin\left(\frac{\pi}{m}\right),
\end{align}
which has two families of solutions: if we call $A$ and $B$ the arguments of the sine functions, then either 
$A=B+2\pi g$ or $A=\pi-B+2\pi g$.
These conditions give rise to the following restrictions on $m$:
\begin{align}
m&=\frac{Q}{g},\qquad g\in\mathbb{N},\quad g<Q,\\
m&=\frac{2Q+2}{2g+1},\qquad g\in\mathbb{N}.
\end{align}
Since $m$ is the number of clusters in the solutions, the only values that have a physical meaning are $m \in \mathbb{N}$. It is important to notice that the solutions provided by this condition reduces the dynamical equation to $\dot{\theta}_l = \omega$, a rigid rotation of constant angular velocity, that does not depend on the coupling parameters. However, such rotation cannot be eliminated by a change of reference frame, as the system is not invariant under rotations.

Finally, let us investigate the second condition for the existence
of clusters. It is sufficient to guarantee that two consecutive
clusters have identical values of the sine argument
\begin{align}
\sin\left(
2\gamma+
\frac{4\pi(l-1)}{m}
+\beta
\right)
=
\sin\left(
2\gamma+
\frac{4\pi l}{m}
+\beta
\right),
\end{align}
which lead to $m=2$ or $4$, independently of the value of $Q$.
In these cases, all clusters have the same non-trivial dynamical
equation, as follow

\begin{align}
m=2\ \mathrm{and}\ Q\ \mathrm{odd}
\quad&\longrightarrow\quad
\dot{\theta} =\omega - \frac{K}{Q}\sin\alpha
- \frac{J}{Q}\sin(2\gamma+\beta), \label{39}\\
m=2\ \mathrm{and}\ Q\ \mathrm{even}
\quad &\longrightarrow\quad \dot{\theta}=\omega. \label{40} \\
m = 4 \quad &\longrightarrow \quad \dot{\theta}=\omega+\frac{K\sin\alpha}{Q}F(Q,4). \label{41}
\end{align}

So, depending on the values of $Q$, we may have active solutions,
in which all oscillators rotate with the same frequency, or static
solutions, in which the oscillators lock into some phase, the
equilibrium locked phases depending on the coupling parameters.
To calculate the case $m=2$ and $Q$ odd equilibrium solutions, we proceed as the $m=1$ case. First, we notice that static real solutions exist only for $|J|\geq|Q\omega-K\sin\alpha|$. Equation (\ref{40}) leads to these families of solutions
\begin{align}
\thx_1&=\frac{1}{2}\left[\arcsin\left(\frac{Q\omega-K\sin\alpha}{J}
\right)-\beta\right]+n\pi,\\
\thx_2&=\frac{1}{2}\left[\pi-\arcsin\left(
\frac{Q\omega-K\sin\alpha}{J}\right)-\beta\right]+n\pi
\end{align}
with $n=0, 1$ to keep the phases in the interval $(0,2\pi)$.

Their linear stability analysis follows the same steps as the
$m=1$ case. The resulting eigenvalues can be written as
\begin{align}
\lambda_p
={}&
\frac{
K\cos\alpha-J\cos(2\gamma+\beta)
}{Q}
\nonumber\\
&+
\frac{
K\cos\alpha+J\cos(2\gamma+\beta)
}{Q}
\sum_{n=1}^{Q}
(-1)^n
\cos\left(\frac{2\pi pn}{N}\right).
\end{align}

In particular, the homogeneous mode $p=0$ for odd $Q$ has eigenvalue
\begin{align}
\lambda_0 =-\frac{2J}{Q}\cos(2\gamma+\beta),
\end{align}
leading to $\thx_1$ being stable and $\thx_2$ unstable for $J > 0$. For negative $J$, they change their stabilities.

Using the stationary-state condition, the eigenvalues can also be written in function of the coupling parameters as
\begin{align}
\lambda_0^\pm=\mp\frac{2}{Q}\sqrt{J^2-\left(Q\omega-K\sin\alpha\right)^2}.
\end{align}

\begin{figure}[h]
    \centering
    \includegraphics[width=0.45\linewidth]{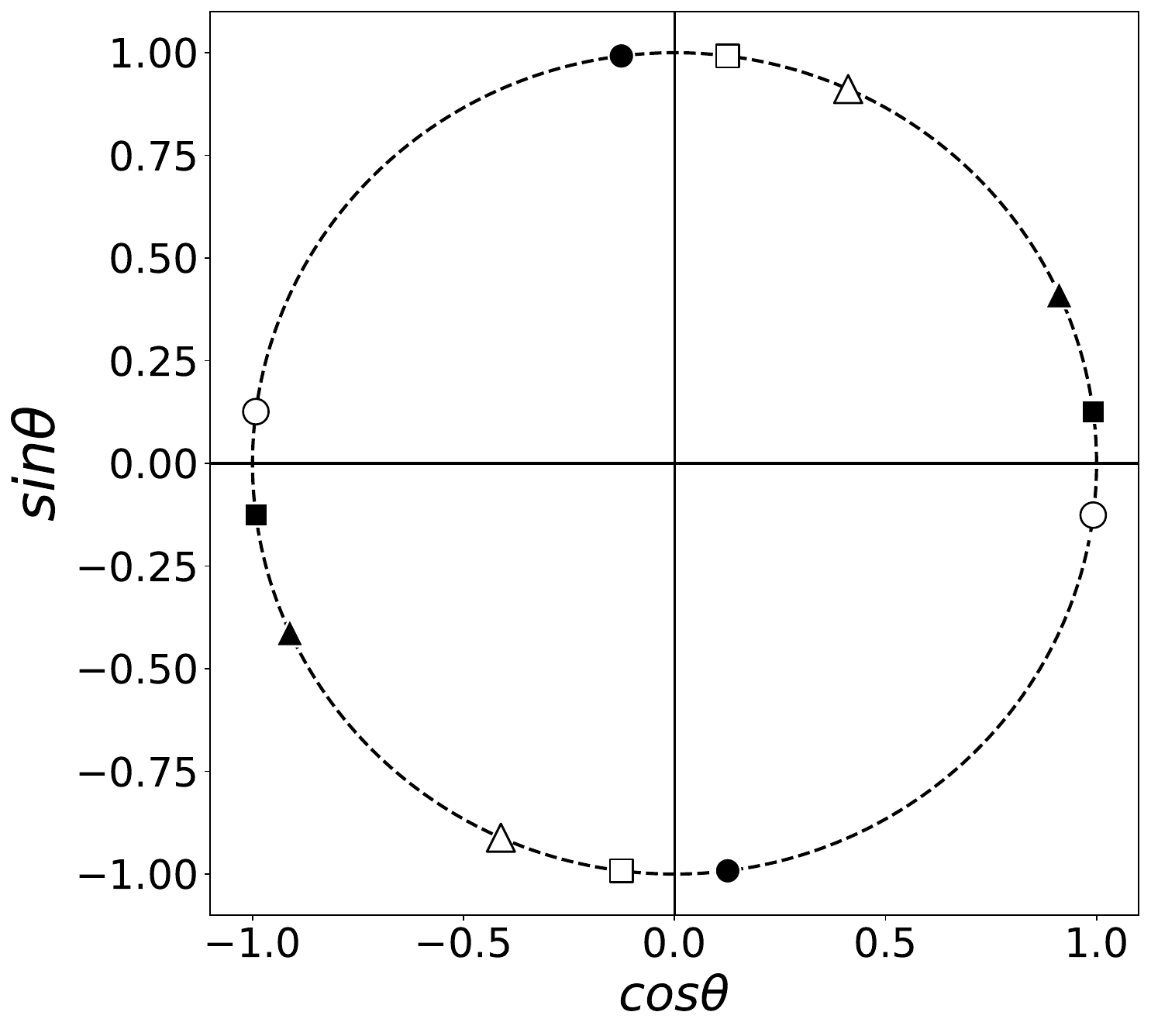}
    \caption{Equilibrium points for $K = \omega = 1$,  $J = 4$ and $Q=1$ (circles and squares) and $3$ (circles and triangles). Full symbols indicates stable fixed points, while empty markers are unstable ones.}
    \label{fig:fixedPoints}
\end{figure}

As examples, we calculated the equilibrium phases for $Q = 1$ and $3$, fixing $\omega = 1$, $K = 1$ and $J = 4$. Eq. \eqref{eq::homo} is independent of $Q$ and, since $|J| \geq \omega$, it generates four equilibrium points, indicated as the circles in Figure \ref{fig:fixedPoints}-(a). For all cases, the empty symbols indicate unstable phases while full ones implies in stable phases. Each of the stable solutions has an attraction basin comprising approximately half of the circumference. The $2-$ cluster solutions are distinct for cases $Q=1$ and $Q=3$ due to the explicit dependence of Eq. $(58)$ on $Q$. Squares on Figure \ref{fig:fixedPoints}-(a) indicate the phases for $Q=1$ while triangles refer to $Q=3$ solutions. 

\subsection{Sequential Ordering - $2Q$ neighbors}

We now consider solutions where oscillators are still divided into $m$ synchronized clusters, but each cluster is composed of consecutive oscillators. Let $L = \dfrac{N}{m}$ be the number of oscillators per cluster and $\gamma_\ell = \gamma + \dfrac{2\pi(\ell-1)}{m}$ be the phase of oscillators in cluster $l$. Under the $L >> Q$ hypothesis, we have two kinds of oscillators: those whose nearest neighbors are within their own cluster, namely bulk oscillators, and those for which at least one nearest neighbor is located on other clusters, termed interface oscillators. 

For the bulk oscillators of a cluster $\ell$, all phases are the same, reducing the dynamics to a similar structure as Eq. \eqref{eq::homo}.
\begin{equation}
\dot{\theta}_{\text{bulk},\ell} = \omega + K\sin \alpha +J\sin\left(  2 \gamma + \beta + \dfrac{4\pi (\ell-1)}{m}\right).
\label{eq::condBulk}
\end{equation}
Therefore, for the existence of an equilibrium solution, all clusters must have the same velocity, implying that the sine term to be $\ell-$independent, leading to $m = 2$ and $4$ as possible solutions, as in the alternate ordered case. In addition, we need to verify the conditions imposed by the interface oscillators. Let us consider the interface between clusters $\ell$ and $\ell+1$. We define $\Delta = \gamma_{\ell+1} - \gamma_{\ell} = \dfrac{2\pi}{m}$. For a typical oscillator in $\ell$, the $s$ right neighbors are within $\ell$ and $Q-s$ belongs to $\ell+1$ cluster. For this oscillator, the dynamical equation is

\begin{align}
\dot{\theta}_{\ell,s}&=\omega+\frac{K}{2Q}\left[(Q+s)\sin\alpha+(Q-s)\sin(\Delta+\alpha)\right]+ \nonumber \\
&+\frac{J}{2Q}\left[(Q+s)\sin(2\gamma_\ell+\beta)+(Q-s)\sin(2\gamma_\ell+\beta+\Delta)
\right].
\end{align}

The whole system being in a equilibrium solution implies that bulk and interface equations to be equivalent. Thus, we need to calculate the conditions leading to $\dot{\theta}_{\text{bulk},\ell} - \dot{\theta}_{\ell,s} = 0$ for all $s$ and $\ell$:

\begin{align}
\dot{\theta}_{\ell,s}-\dot{\theta}_{\mathrm{bulk},\ell}
=\frac{Q-s}{Q}\sin\left(\frac{\pi}{m}\right)
\Bigg[K\cos\left(\alpha+\frac{\pi}{m}\right)
+J\cos\left(2\gamma_\ell+\beta+\frac{\pi}{m}
\right)\Bigg],
\end{align}

leading to the following condition 
\begin{equation}
K\cos\left(\alpha+\frac{\pi}{m}\right)+J\cos\left(
2\gamma_\ell+\beta+\frac{\pi}{m}\right)=0.
\label{eq::condInt}
\end{equation}

Since we have additional conditions due to Eq. \eqref{eq::condBulk}, we can test $m = 2$ or $4$ to see if these values satisfy Eq. \eqref{eq::condInt}. 

For the $m=2$ case, the condition obtained from the bulk oscillators is

\begin{equation}
\sin(2\gamma_\ell+\beta) = -\frac{\omega+K\sin\alpha}{J},
\end{equation}
whereas the interface condition given by Eq. (50) reduces to
\begin{equation}
\sin(2\gamma_\ell+\beta) = -\frac{K\sin\alpha}{J}.
\end{equation}
Therefore, requiring both the bulk and interface oscillators to have the same stationary velocity imposes $\omega=0$. Hence, for nonzero natural frequency, the $m=2$ bulk-ordered configuration cannot form a fully stationary equilibrium satisfying simultaneously the bulk and interface conditions. This contrasts with the sequential-ordering case, where the corresponding cluster phases can satisfy the equilibrium condition for nonzero $\omega$. Finally, for the $m=4$ case the bulk and interface conditions cannot be simultaneously satisfied for $J \neq 0$. When $J=0$, the solution exist provided that the frustration parameter $\alpha = \pi/4$ or $ 5\pi/4$.

\section{GENERIC SOLUTIONS}
\label{sec:general}

\begin{figure}[h]
    \centering
    \includegraphics[width=0.46\linewidth]{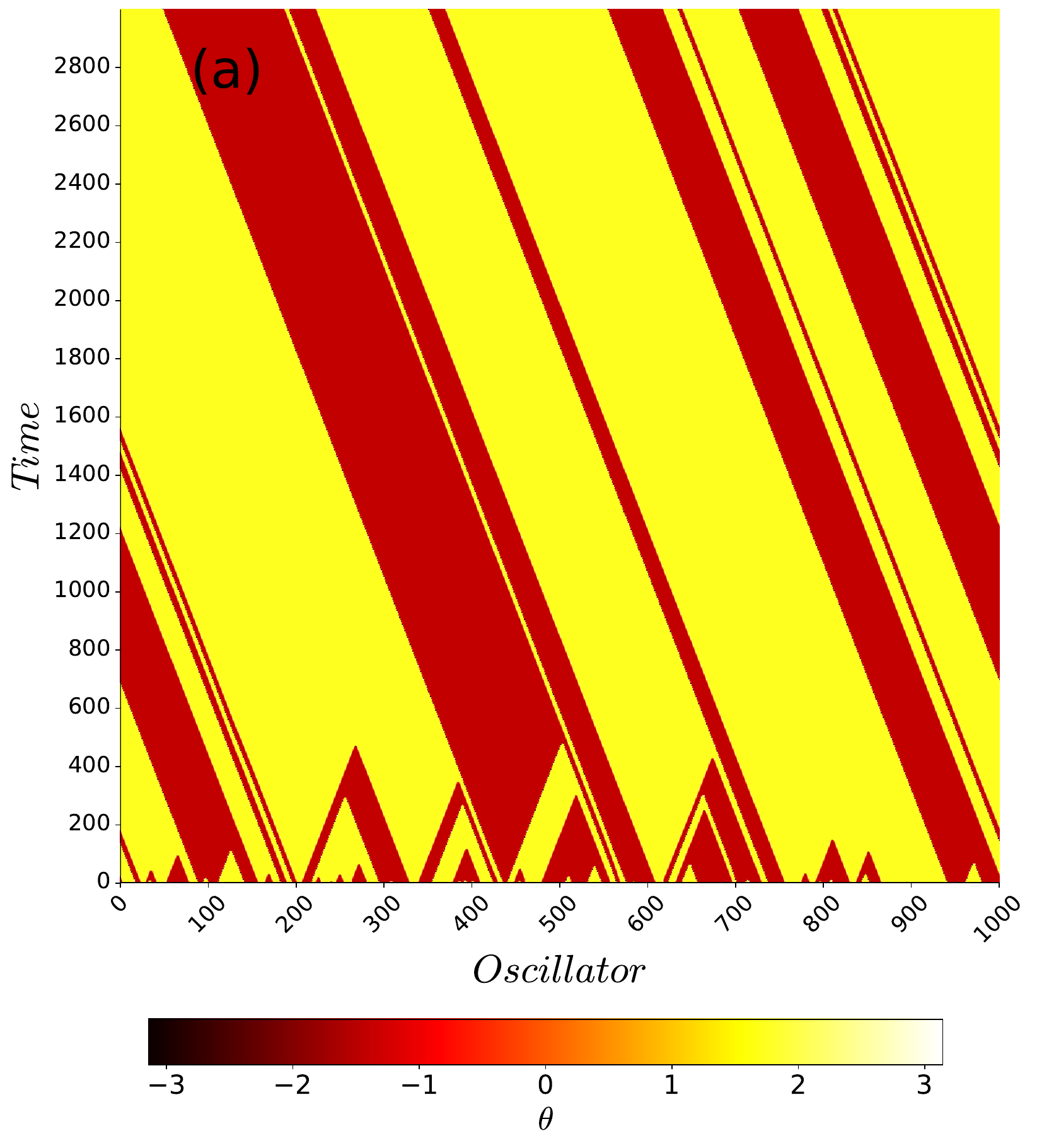}
    \includegraphics[width=0.51\linewidth]{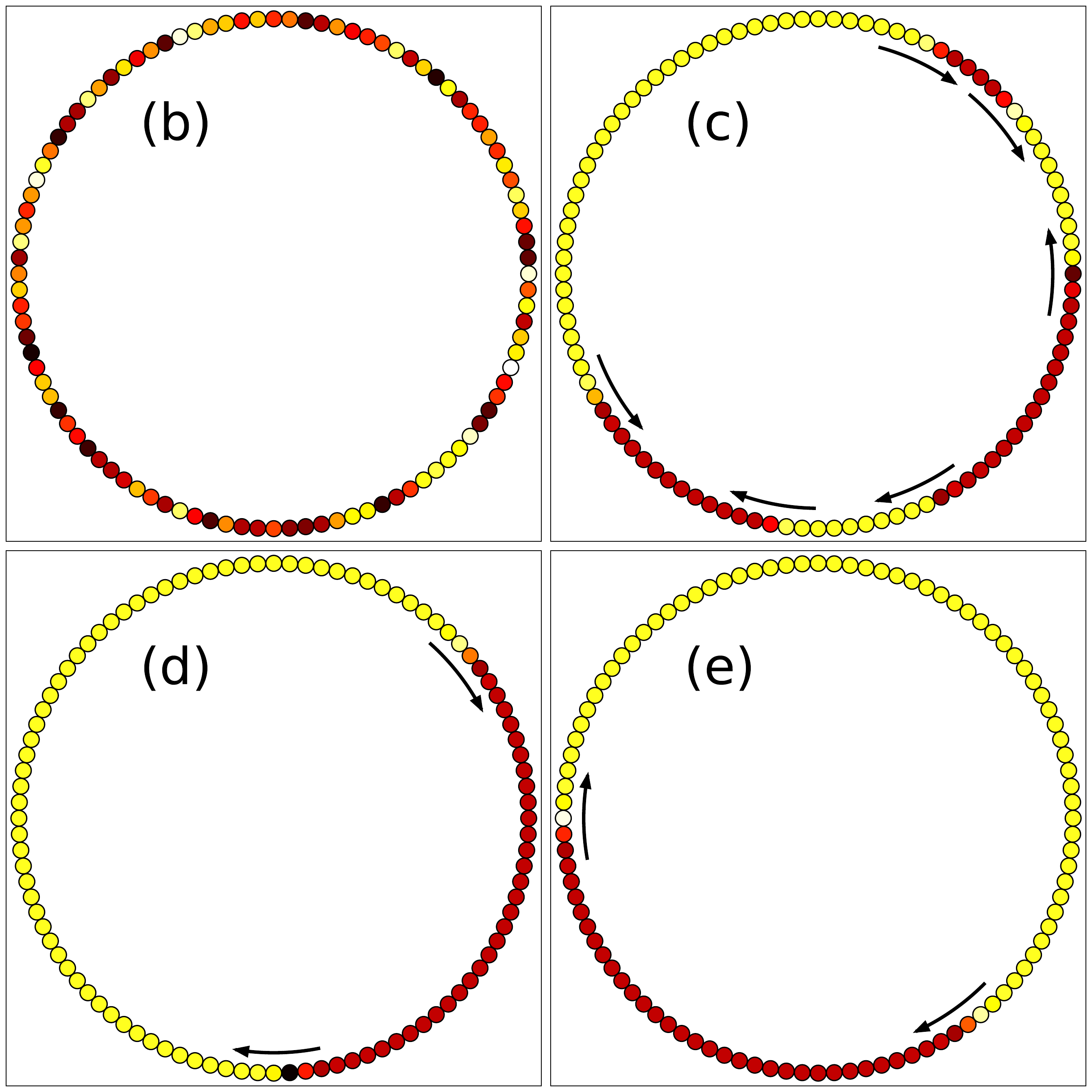}
    \caption{Heatmap of oscillators' phases of a typical simulation of the matrix-coupled Kuramoto model on a ring-like structure. (b) to (e) Snapshots of the individual phases showing the temporal evolution of the clusters. The color coding is the same for the heatmap and the right figures. See supplemental videos in \cite{codes}.}
    \label{fig::typical}
\end{figure}

The specific configurations discussed in the previous sections do not exhaust all types of solutions that exist in this system. In Figure \ref{fig::typical}-(a), we show the heatmap of a typical simulation with $N=1000$ oscillators and initial phases draw from a uniform distribution $U(0,2\pi)$. It can be seen that the oscillators converge initially to one of the stable solutions of Eq. $(3)$, the specific arrangement depending on the initial condition. However, this state is not a static solution of the system, since oscillators at the interface of a cluster drift to the next cluster, creating a juggling mechanism where clusters exchange particles. The pattern is similar to a "traveling wave" dynamics on the interface between clusters. As can be seen from figure \ref{fig::typical}-(a), all interfaces move with the same speed. If two neighboring clusters drift in opposite directions, they may collide and annihilate each other. When all resulting interfaces move in the same direction the system reaches an active stationary state, in which the size of the clusters does not change, just the oscillators belonging to them. Figures \ref{fig::typical}-(b) to (e) show an example of the convergence of this clusters patterns. Starting from a random initial condition at \ref{fig::typical}-(b), the system arranges itself in several clusters, \ref{fig::typical}-(c), in which their interfaces drifts on a given direction, indicated by the arrows. If the two interfaces of a cluster start drift inwards, this cluster will be engulfed by the neighboring ones, as can be seen in \ref{fig::typical}-(d) and (e). From now on, the interfaces just travel as a stationary wave. A video showing the full time evolution of the clusters can be found in the github repository \cite{codes}
.

\begin{figure}[h]
    \centering
    \includegraphics[width=0.45\linewidth]{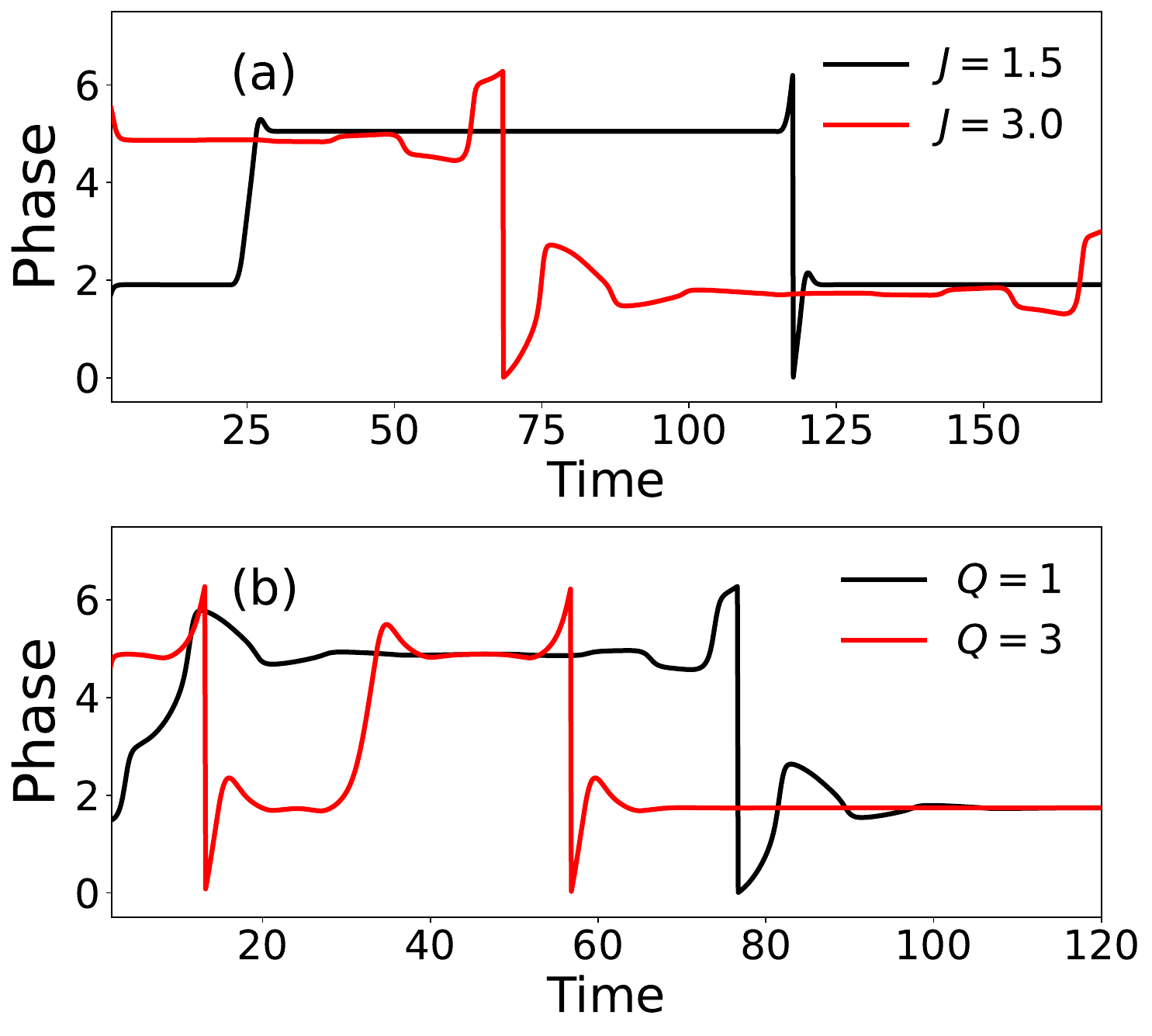}
    \includegraphics[width=0.45\linewidth]{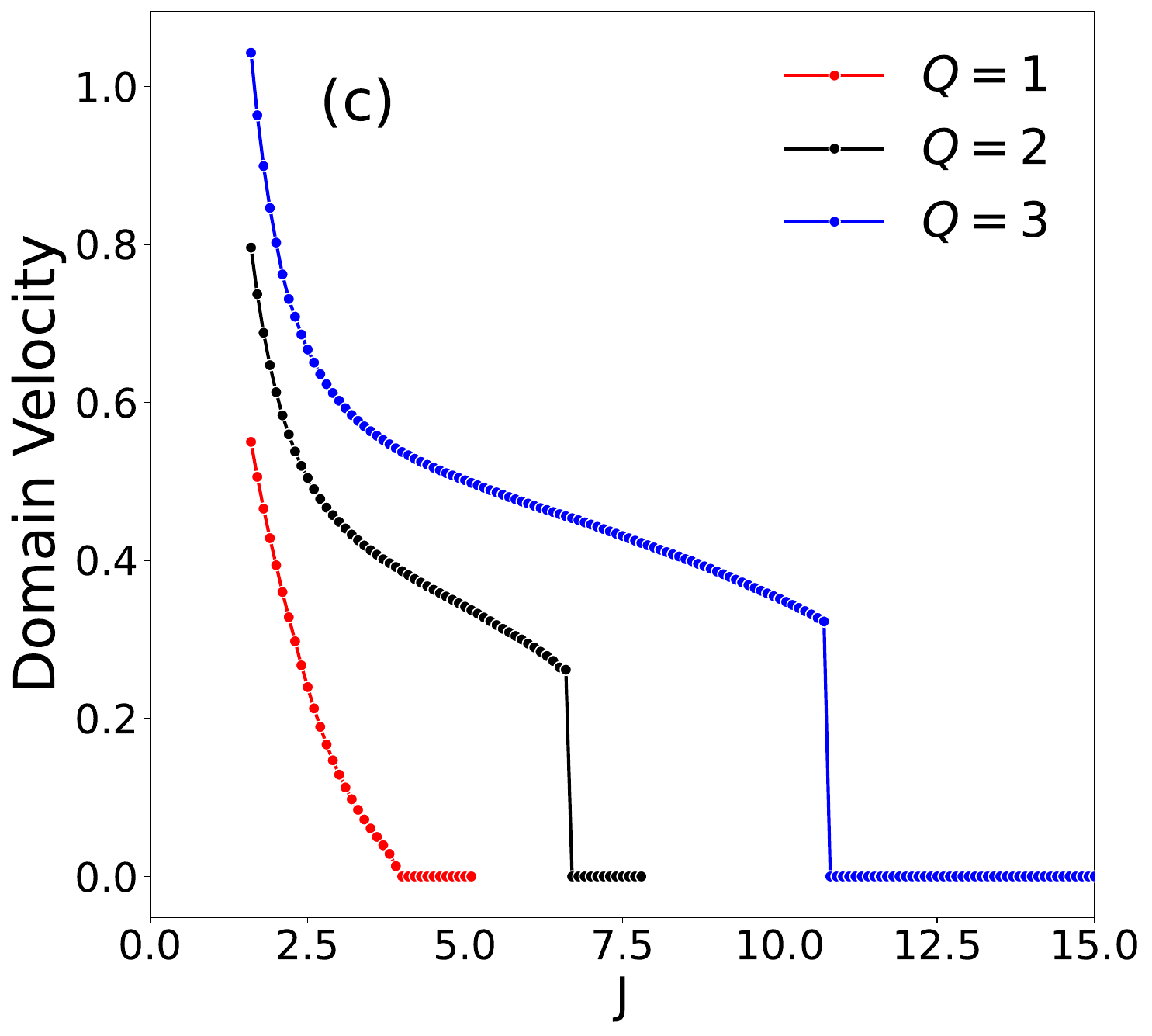}
    \caption{Phase evolution of typical oscillators: (a) $Q=1$ for $J=1.5$ and $J=3$ and (b) $J=3.5$ for $Q=1$ and $Q=3$; (c) Domain angular velocity as a function of $J$ for some values of $Q$.}
    \label{fig::phaseOscillation}
\end{figure}

Let us investigate what happens on the vicinity of a interface. Figures \ref{fig::phaseOscillation}-(a) and (b) shows a typical trajectory for a single oscillator, i.e., a vertical slice of the heatmap of Figure \ref{fig::typical}-(a). It can be noticed that the oscillators maintain its phase fixed for a certain amount of time, related to the width of the fringes, and then start drifting from it in an increasing amplitude motion, dragged for the antipodal equilibrium point, in which the oscillatory motion fades. The distinct trajectories of those figures, associated to distinct coupling parameters, indicate the dependency of this dragging force with $J$ and $Q$, being more aggressive the larger the $J$
and smaller the $Q$.

In addition, we also measured the absolute domain velocity $\vartheta$ by using the slope of the fringes in the heatmaps. The values are compiled in Figure \ref{fig::phaseOscillation}-(c) for several values of $J$ and $Q$. In general, we observe the same pattern independent of the number of neighboring links. Until a certain threshold, the larger the $J$, the smaller the $\vartheta$, indicating that stronger couplings tend to maintain the oscillators within their clusters for longer, effectively phase locking them if $J$ is large enough, preventing the juggling mechanism. Regarding the influences of $Q$, for fixed $J$, the inclusion of more connections favors the interface dynamics, increasing $\vartheta$. In addition, for $Q>1$, the transition from dynamic to static interfaces changes its nature, becoming discontinuous.

\begin{figure}[h]
    \centering
    \includegraphics[width=0.9\linewidth]{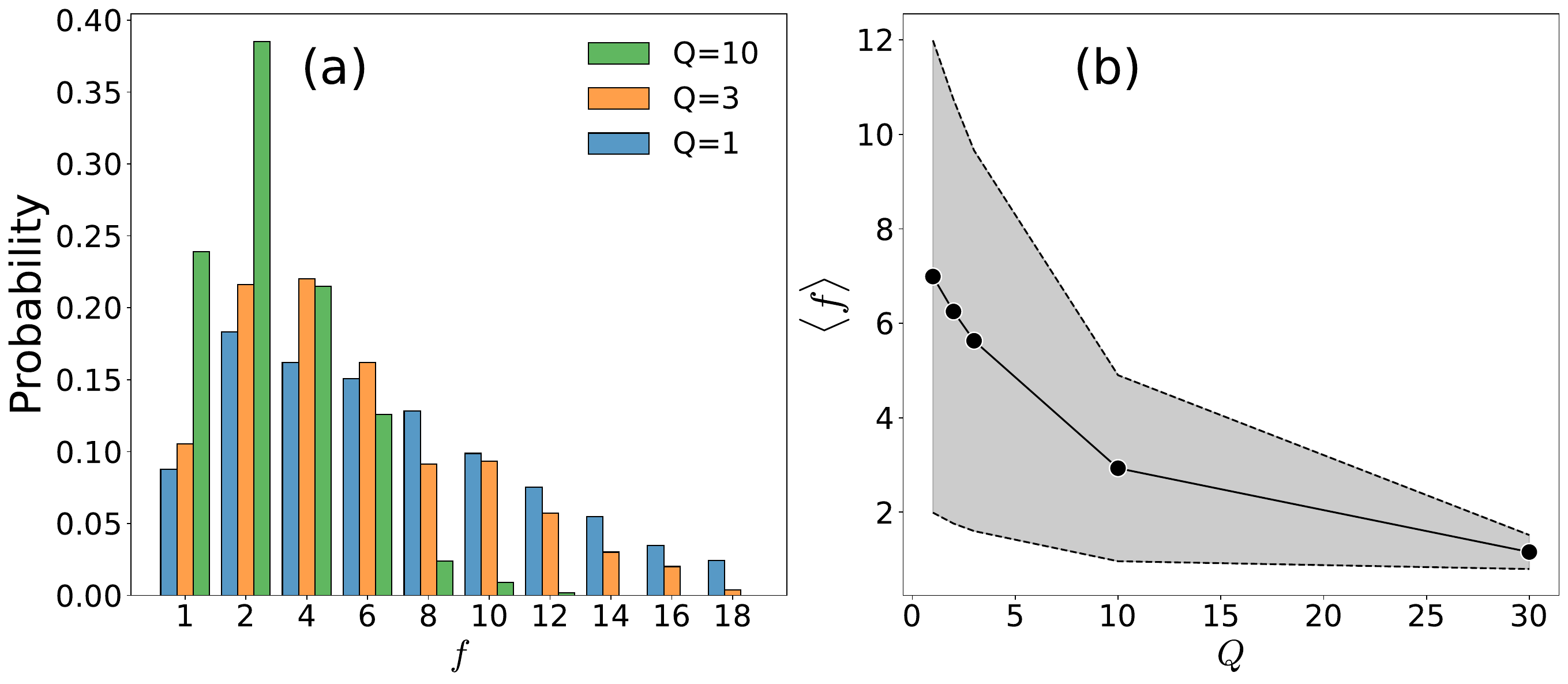}
    \caption{(a) Probability of finding the system with $f$ stationary fringes, computed from $1000$ simulations of the model for some values of $Q$. (b) Average number of fringes $\av{f}$ as a function of $Q$. The shaded region indicates the standard deviation of the distributions.}
    \label{fig::fringeDistribution}
\end{figure}

Other interesting metric of this general solutions is to see the distribution of the fringes from distinct random initial conditions. Since this dynamic admit only two major stable collective states, and the synchronized domains necessarily alternate between them, traversing the ring requires every transition from one state to the other to be followed by a transition back to the original state before completing the loop. Therefore, the number of distinct domains must always be even, with the only exception being the fully synchronized state, in which the entire system occupies a single domain. Figure \ref{fig::fringeDistribution}-(a) shows the distribution of fringes $f$ for $1000$ independent simulations, for some cases of $Q$. It can be seen that they are asymmetrical distributions peaked on $f=2$ with decreasing in the dispersion as $Q$ increases. For $Q=30$ (data not shown) the probability of the system to attain full synchronization reaches up to $80\%$.  In figure \ref{fig::fringeDistribution}-(b) we show for more values of $Q$ the average number of fringes and one standard deviation, extracted for similar distributions as Figure \ref{fig::fringeDistribution}-(a). As expected, as $Q$ increases, both $\av{f}$ and $\sigma f$ decreases asymptotically to $1$ and $0$, respectively.

\section{FINAL REMARKS}
\label{sec:discussion}
In this work, we investigated the dynamics of matrix-coupled identical Kuramoto oscillators arranged in ring networks, focusing on the effects of the interaction range and the generalized frustration introduced by the coupling matrix. By considering interactions with up to $2Q$ neighbors, we analyzed the asymptotic states obtained from generic initial conditions and constructed particular solutions for multiple synchronized clusters. 

Our analytical results show that the ring topology supports several families of clustered solutions, depending on the arrangement of the particles. For alternate ordered oscillators, we have found two classes of solutions. The first consists of one or two static clusters, that lock on certain phases, depending on the coupling parameters and despite the natural oscillation frequency $\omega \neq 0$. The second one is comprised of clusters that rotate rigidly with frequency $\omega$. The allowed quantity of clusters depends on the neighbor connectivity $Q$. We also investigated sequential ordered clusters and showed that the constraints imposed by the interface oscillators further restrict the possible equilibrium configurations.

The numerical simulations reveal a different type of collective behavior when the system is initialized with random phases. Instead of converging to one of the static cluster configurations considered analytically, the oscillators organize into dynamical clusters whose interfaces continuously move along the ring. The oscillators within each cluster remain close to one of the stable phase configurations, while particles are exchanged at the interfaces. The resulting pattern therefore preserves the cluster structure while the individual oscillators belonging to each cluster change with time, producing a dynamics analogous to a juggling process. When interfaces moving in opposite directions collide, the corresponding clusters annihilate, until a state with interfaces moving in a common direction is reached.

The velocity of these dynamical interfaces depends on both the coupling strength $J$ and the interaction range $Q$. For fixed $Q$, an increase in $J$ reduces the domain velocity until the interfaces become static, indicating that sufficiently strong frustration can suppress the juggling dynamics. Conversely, increasing the number of interacting neighbors enhances the interface motion. For $Q > 1$,  the transition between dynamical and static interfaces is  discontinuous, showing that the interaction range does not merely change the characteristic velocity of the interfaces but also affects the nature of the transition between the two regimes.

As $Q$ increases, the distribution of the number of domains becomes increasingly concentrated around the fully synchronized configuration. In particular, the average number of domains decreases toward one, while its dispersion decreases, indicating that increasing the interaction range favors global synchronization over fragmented dynamical states.

These results show that generalized frustration combined with local ring connectivity can produce collective states that are qualitatively different from both conventional synchronization and static cluster solutions. In particular, the juggling states provide a mechanism through which the spatial organization of the oscillators remains stationary at the macroscopic level while the individual oscillators continuously migrate between clusters. The coexistence of analytically tractable cluster solutions and dynamically evolving interfaces makes the matrix-coupled ring a useful setting for studying the interplay between frustration, topology, and multistability.

\begin{acknowledgements}
 This work was partly supported by FAPESP [Grant 2025/08895-4 (G.S.C.)] and [Grant 2021/ 14335-0 (M.A.M.A.)] and by CNPq, Brazil, [Grant 303814/2023-3 (M.A.M.A.)]. We thank the Coaraci Supercomputer for computer time (FAPESP grant 2019/17874-0) and the Center for Computing in Engineering and Sciences at Unicamp (FAPESP grant 2013/08293-7).
\end{acknowledgements}

\end{document}